\documentclass[iop,numberedappendix]{emulateapj}
\usepackage{graphicx}
\usepackage{natbib}
\usepackage{amsmath}
\newcommand{\rff}[1]{Fig.\ \ref{#1}}

\begin{document}
\title{Inducing chaos by breaking axial symmetry in a black hole magnetosphere}
\shorttitle{Inducing chaos by breaking axial symmetry\ldots}

\author{O. Kop\'{a}\v{c}ek and V. Karas}
\affil{Astronomical Institute, Academy of Sciences, Bo\v{c}n\'{i} II, CZ-141\,31~Prague, Czech~Republic}
\email{kopacek@ig.cas.cz}

\begin{abstract}
While the motion of particles near a rotating, electrically-neutral (Kerr), and charged (Kerr--Newman) black hole is always strictly regular, a perturbation in the gravitational or the electromagnetic field generally leads to chaos. The transition from regular to chaotic dynamics is relatively gradual if the system preserves axial symmetry, whereas non-axisymmetry induces chaos more efficiently. Here we study the development of chaos in an oblique (electro-vacuum) magnetosphere of a magnetized black hole. Besides the strong gravity of the massive source represented by the Kerr metric we consider the presence of a weak, ordered, large-scale magnetic field. An axially-symmetric model consisting of a rotating black hole embedded in an aligned magnetic field is generalized by allowing an oblique direction of the field having a general inclination, with respect to the rotation axis of the system. The inclination of the field acts as an additional perturbation to the motion of charged particles as it breaks the axial symmetry of the system and cancels the related integral of motion. The axial component of angular momentum is no longer conserved and the resulting system thus has three degrees of freedom. Our primary concern within this contribution is to find out how sensitive the system of bound particles is to the inclination of the field. We employ the method of the maximal Lyapunov exponent to distinguish between regular and chaotic orbits and to quantify their chaoticity. We find that even a small misalignment induces chaotic motion.
\end{abstract}


\section{INTRODUCTION}
\label{intro}
Since the seminal work of \citet{carter68}, it has been well-known that the motion of particles near a rotating black hole is strictly regular. This remarkable property is guaranteed by the full integrability of the system due to the existence of the fourth constant of motion, and it holds even in the case of an electrically-charged rotating black hole. On the other hand, as soon as the black hole is embedded in an external magnetic field, chaotic dynamics may appear \citep{kopacek10}. While the electric charge of astronomical black holes is negligible, the role of the magnetic field is important \citep[e.g.,][]{camendzin07}.

In this paper we study the role of large-scale (organized) magnetic fields on the properties of motion near a rotating black hole. Recent general-relativistic magnetohydrodynamic (GRMHD) simulations \citep{penna10} suggest that accretion flows onto black holes are complemented by outflows emerging from accretion disks and tori, and, in some regions of the flow, produce large-scale (organized) magnetic loops in a self-consistent manner. The mechanism leading to the development of ordered bundles of the magnetic lines of force with a significant degree of long-range coherence appears to be highly relevant in the context of formation of jets and outflows from accreting black holes. Simulations indicate that the role of black hole rotation is important and this supports the idea of Blandford--Znajek mechanism \citep{blandford77} as the origin of acceleration and collimation in the vicinity of black holes \citep{sadowski13}. The formulation of the problem of particle acceleration in the context of oblique pulsar magnetospheres has been discussed by \citet{li12}.

We explore the effects of organized magnetic fields in a different context of particle motion near a weakly-magnetized black hole (the magnetic field does not change the spacetime metric). We address the problem of regularity versus chaoticity of the resulting motion of electrically charged particles (electrons and ions) resulting from the mutual interplay of gravitational, electromagnetic, and gravito-magnetic effects of general relativity. As a matter of principle, we want to understand under what condition chaos emerges and drives the particle motion near magnetized black holes. To this end we assume that the magnetic field arises from currents flowing far out in the accretion disk. Unlike most of previous analytical works, we do not impose axial symmetry (the magnetic field can be inclined with respect to the rotation axis of the black hole).

The particle approximation allows us to concentrate our attention on purely general-relativistic effects of the curved electro-vacuum spacetime while neglecting collective interactions and shocks in the plasma \citep{kalapotharakos12}. The adopted approximation is thus relevant to situations when the GRMHD conditions are {\rm not} satisfied: in our case, particle mean free path $l_{\rm f}$ is assumed to be very long (exceeding the gravitational radius of the black hole, $l_{\rm f}>R_{\rm g}\equiv GM/c^2$) while, simultaneously, the gravitational field is very strong (curvature radius\footnote{Curvature of the vacuum spacetime may be locally characterized by Kretschmann scalar $K$ evaluated from the Riemann curvature tensor $K\equiv R^{\mu\nu\xi\pi}R_{\mu\nu\xi\pi}$. Explicit form of Kretschmann scalar for Kerr--Newman spacetime is given by \citet{henry00}. Characteristic length scale of the spacetime time curvature (curvature radius) is then expressed as $K^{-1/4}$.} is smaller than $R_{\rm g}$); see \citet{cremaschini13} and references therein.

This work represents a natural extension of our previous studies \citep{kopacek10,kovar10,kovar08} in which we considered dynamic properties of charged matter orbiting in the vicinity of massive objects. Our primary concern was to identify and investigate regions of a stable, off-equatorial motion. Trajectories of ionized particles confined to such regions constitute a basic {\em non-interacting test particle} approximation of astrophysical coronae formed by diluted plasma above and below the accretion disk of accreting systems. 

Several models have been considered so far. As a first step, the existence of off-equatorial orbits was investigated in the spacetime described by the exact Kerr--Newman solution, i.e., the case of a rotating electrically charged black hole. Considering only the astrophysically-relevant region above the outer horizon of the black hole, it was found that there are no stable circular orbits ({\em halo orbits}) outside the equatorial plane \citep{kovar08}. Therefore, we altered our model, namely, we employed test field solutions describing large-scale, ordered magnetic fields (asymptotically uniform or dipole type) in which the compact object is immersed.  In particular, we investigated a system consisting of Schwarzschild black hole with a rotating magnetic dipole field, and the case of the Kerr black hole in an asymptotically-uniform magnetic field, which both appeared to host non-equatorial confinements of charged particles \citep{kovar10}. Since both of these systems were found to be non-integrable, in which case the chaotic dynamics emerges, we subsequently focused on the dynamic properties of these orbits, trying to identify which parameters of the system trigger chaos. We found that within the off-equatorial lobes, the dynamics are mostly regular and chaos typically appears only when the energy of the particle is sufficiently increased  to the level that corresponds to cross-equatorial confinements \citep{kopacek10}. 

More recently, a static model of an exact Einstein--Maxwell spacetime was considered in this context \citep{kovar13}. Namely, the dynamics of charged matter in off-equatorial wells above the massive magnetic dipole described by Bonnor's exact solution \citep{bonnor66} was investigated. In such a case, the electromagnetic field affects the geometry of the spacetime and as a particular consequence of this influence, we found that the system allows off-equatorial orbits even for neutral test particles. The interplay between the chaotic and ordered motion of test particles in the exactly-given static field of a massive source encircled by a disk or a ring was also recently investigated \citep{sukova13,semerak12,semerak10}. Nevertheless, in the following we intend to focus on a more realistic model in which the rotation of the central object is included. On the other hand, we will restrict ourselves to the test field approximation in which the electromagnetic field affects charged particles, but it does not modify the metric.

In this contribution we plan to investigate the dynamical properties of charged particles in the generalized model consisting of the rotating Kerr black hole in the asymptotically-uniform magnetic field that is inclined with respect to the rotation axis. A corresponding test field solution was given by \citet{bicak85}. To our knowledge, however, the particle motion in this setup has not yet been inspected. 

The paper is organized as follows. In Section\ \ref{formalism} we present formal description of the model and explicitly specify the electromagnetic four-potential $A_\mu$. Then we review the Hamiltonian formalism which is employed to derive equations of motion. Application of the method of the maximal Lyapunov exponent as an indicator of chaos is also briefly discussed. Analysis of the motion of particles affected by the inclination of the magnetic field is presented in Section\ \ref{dynamics}. Three distinct classes of orbits are treated separately in corresponding subsections. Possible ways for observational verification are discussed in Section\ \ref{discussion}. Results of the numerical analysis are summarized and discussed in Section\ \ref{conclusions}. In Appendix\ \ref{appa} we comment on the method of the effective potential and its limited applicability in given context. Appendix\ \ref{appb} provides an estimate of radiation power generated by the investigated system.

\section{SPECIFICATION OF THE MODEL AND EMPLOYED METHODS}
\label{formalism}
\subsection{Rotating Black Hole in External Magnetic Field}
The Kerr metric  describing the geometry of the spacetime around the rotating black hole may be expressed in Boyer--Lindquist coordinates $x^{\mu}= (t,\:r, \:\theta,\:\varphi)$ as follows \citep{mtw}:
\begin{eqnarray}
\label{metric}
{\rm d}s^2&=&-\frac{\Delta}{\Sigma}\Big({\rm d}t-a\sin{\theta}\,{\rm d}\varphi\Big)^2\\& &\nonumber+\frac{\sin^2{\theta}}{\Sigma}\Big[\big(r^2+a^2)\,{\rm d}\varphi-a\,{\rm d}t\Big]^2+\frac{\Sigma}{\Delta}\,{\rm d}r^2+\Sigma\, {\rm d}\theta^2,
\end{eqnarray}
where
\begin{equation}
{\Delta}\equiv{}r^2-2Mr+a^2,\;\;\;
\Sigma\equiv{}r^2+a^2\cos^2\theta.
\end{equation}
We stress that geometrized units are used throughout the paper. Values of basic constants (gravitational constant $G$, speed of light $c$, Boltzmann constant $k$, and Coulomb constant $k_C$) thus equal unity $G=c=k=k_C=1$.

Test field solution corresponding to the aligned magnetic field (of the asymptotic strength $B_z$) was derived by \citet{wald74}. This solution was later generalized by \citet{bicak85} to describe the field which is arbitrarily  inclined with respect to the rotation axis (specified by two independent components $B_z$ and $B_x$). Here we also consider a non-zero electric charge $Q$ of the black hole generating the electromagnetic field of the Kerr--Newman black hole, though in the test field regime (metric remains unaltered by $Q$). The resulting vector potential $A_{\mu}$ is given as follows:  

\begin{eqnarray}
\label{empot1}
A_t&=&\frac{B_{z}aMr}{\Sigma}\left(1+\cos^2\theta\right)-B_{z}a\\& &\nonumber+\frac{B_xaM\sin\theta\cos\theta}{\Sigma}\left(r\cos\psi-a\sin\psi\right)-\frac{Qr}{\Sigma}\\
A_r&=&-B_x(r-M)\cos\theta\sin\theta\sin\psi\label{empot2}\\
A_{\theta}&=&-B_xa(r\sin^2\theta+M\cos^2\theta)\cos\psi\\& &\nonumber-B_x(r^2\cos^2\theta-Mr\cos2\theta+a^2\cos2\theta)\sin\psi\label{empot3}\\
A_{\varphi}&=&B_z\sin^2\theta\left[\frac{1}{2}(r^2+a^2)-\frac{a^2Mr}{\Sigma}(1+\cos^2\theta)\right]\label{empot4}\\& &\nonumber-B_x\sin\theta\cos\theta\Big[\Delta\cos\psi\\& &\nonumber+\frac{(r^2+a^2)M}{\Sigma}\left(r\cos\psi-a\sin\psi\right)\Big]+\frac{Qra\sin^2\theta{}}{\Sigma},
\end{eqnarray}
where we use the azimuthal coordinate $\psi$ of the Kerr ingoing coordinates, which is related to Boyer--Lindquist coordinates as follows:
\begin{equation}
\label{kicpsi}
\psi=\varphi+\frac{a}{r_{+}-r_{-}}\ln{\frac{r-r_{+}}{r-r_{-}}},
\end{equation}
with $r_{\pm}\equiv M \pm\sqrt{M^2-a^2}$ denoting the outer and the inner horizon, respectively. We notice that $\lim_{r\to \infty}\psi=\varphi$.

We have previously investigated the structure of the electromagnetic field emerging in a more general case of the Kerr black hole immersed in the asymptotically-uniform magnetic field (1) which is inclined with respect to the rotation axis and (2) in which the black hole is drifting with the constant velocity and in the arbitrary direction \citep{karas12,karas09}. We found complex structures of both the magnetic and the electric field lines in this model. In particular, we observed that due to combined effects of the translational motion and the frame dragging (caused by the black hole's rotation), the null points of magnetic field may arise. These may be highly relevant for the acceleration processes of ionized matter in the vicinity of an accreting object. Nevertheless, the motion of charged particles exposed to such a field has not been inspected in the aforementioned studies and within this paper we consider the inclination of the magnetic field but not the translational motion of the black hole.

\subsection{Equations of Motion}
Employing the Hamiltonian formalism, we first construct the super-Hamiltonian $\mathcal{H}$:
\begin{equation}
\label{hamiltonian}
\mathcal{H}=\textstyle{\frac{1}{2}}g^{\mu\nu}(\pi_{\mu}-qA_{\mu})(\pi_{\nu}-qA_{\nu}),
\end{equation}
where $q$ is the charge of the test particle (of rest mass $m$), $\pi_{\mu}$ is the generalized (canonical) momentum, $g^{\mu\nu}$ is the metric tensor, and $A_{\mu}$ denotes the vector potential of the electromagnetic field. The latter is related to the electromagnetic tensor $F_{\mu\nu}$ by $F_{\mu\nu}=A_{\nu,\mu}-A_{\mu,\nu}$. 

Hamilton's equations of motion are given as
\begin{equation}
\label{hameq}
\frac{dx^{\mu}}{d\lambda}\equiv p^{\mu}=
\frac{\partial \mathcal{H}}{\partial \pi_{\mu}},
\quad 
\frac{d\pi_{\mu}}{d\lambda}=-\frac{\partial\mathcal{H}}{\partial x^{\mu}},
\end{equation}
where $\lambda=\tau/m$ is the affine parameter (dimensionless in geometrized units), $\tau$ denotes the
proper time, and $p^{\mu}$ is the standard kinematical four-momentum for
which the first equation reads $p^{\mu}=\pi^{\mu}-qA^{\mu}$, and thus the conserved value of super-Hamiltonian is equal to $\mathcal{H}=-m^2/2$.
Moreover, the system is stationary since the Hamiltonian is independent on the coordinate time $t$. Its conjugate momentum $\pi_t$ is therefore integral of motion. Namely, it expresses (negatively taken) the energy of the test particle $\pi_t\equiv-E$.

The mass of the black hole $M$ is used to scale all quantities, which is formally equivalent to setting $M=1$ in the equations. We also switch to specific quantities $q/m\rightarrow q$ and $E/m\rightarrow E$ when describing the test particle. From the above equations, we conclude that independent parameters of the investigated system are thus $a$, $qQ$, $qB_z$, and $qB_x$. Indeed, the quantities $B_x$, $B_z$, and $Q$ only appear in the product with the particle's charge in the equations of motion (\ref{hameq}) since the electromagnetic field is treated in the test field approximation. Besides that, particular trajectory is further specified by its initial position in the phase space $r(0)$, $\theta(0)$, $\varphi(0)$, $\pi_r(0)$, $\pi_\theta(0)$, and $\pi_\varphi(0)$ whose values are, however, bound by the normalization condition $p^{\mu}p_\mu=-m^2$. Therefore, we do not set all the components arbitrarily and choose to compute the value $\pi_\theta(0)$ using the normalization of the four-momentum and its remaining components instead.

\subsection{Maximal Lyapunov Characteristic Exponent}
Maximal Lyapunov exponent $\chi$ is commonly used as a basic quantitative indicator of chaotic dynamics \citep[e.g.,][]{lieberman}. Its value directly captures the tendency of nearby orbits to diverge as the system evolves. In other words, it allows us to express how sensitive the given orbit is on the initial condition whilst the high (exponential) sensitivity is a defining property of chaos. The exponent $\chi$ is defined as follows\footnote{The Lyapunov exponent $\chi$ is defined as an asymptotic measure (\ref{mle}). In the numerical application we actually  compute the quantity usually denoted as {\em finite time Lyapunov exponent}, which depends on the integration variable, instead of the limit. However, in this text we disregard such distinctions and use the term Lyapunov exponent for both quantities as it cannot cause any confusion within the scope of the paper.}

\begin{equation}
 \label{mle}
\chi\equiv\lim_{\lambda\to\infty}\frac{1}{\lambda}\ln\frac{||w(\lambda)||}{||w(0)||},
\end{equation}
where we choose to use the standard $L^2$ (Euclidean) norm to measure the length of the deviation vector in the phase space $w(\lambda)=(\delta t,\delta r, \delta \theta, \delta \varphi, \delta\pi_t, \delta\pi_r, \delta\pi_\theta, \delta\pi_\varphi)$. In this context, we have also experimented with norms reflecting the curvature of the spacetime, however, it appeared that for a given application the choice of the norm is not crucial \citep{kopacek10}. Although Lyapunov exponents within the relativistic framework are generally not invariant under coordinate transformations \citep{karas92,dettmann95}, signs of the exponents are preserved \citep{motter03,motter09}. Therefore, the distinction between chaotic and regular orbits may be drawn invariantly.

The usual method of determining the evolution of the deviation $w(\lambda)$ consists in solving variational equations which restricts us to the linear term in a corresponding Taylor expansion \citep{kaltchev13}. One can therefore express the variational equations in the following matrix form with column vector $w^{\rm{T}}$ and its derivative $\dot{w}^{\rm{T}}$ with respect to the affine parameter $\lambda$: 
\begin{equation}
 \label{vareq}
\dot{w}^{\rm{T}}=J\cdot D^2\mathcal{H}(\lambda)\cdot w^{\rm{T}},
\end{equation}
 where $D^2\mathcal{H}$ is the Hessian matrix composed of second derivatives of the super-Hamiltonian $\mathcal{H}$ with respect to phase space variables and $J$ is constant matrix with following block form: $J=\left( \begin{smallmatrix}0_n&I_n\\-I_n&0_n \end{smallmatrix} \right)$, in which $I_n$ and $0_n$ denote $n$-dimensional identity and zero matrices, respectively. In our case $n=4$. Detailed derivations of variational equations (\ref{vareq}) and their properties are given in a recent review on Lyapunov exponents by \citet{skokos10}. Matrix $D^2\mathcal{H}$ is evaluated along the orbit and explicitly depends on the current position in the phase space; therefore it is necessary to integrate variational equations simultaneously with corresponding equations of motion (\ref{hameq}). We set the initial deviation as follows $w(0)=1/\sqrt{8}\left(1,1,1,1,1,1,1,1\right)$. The theory of Lyapunov spectra guarantees that setting random initial deviation results in the computation of the maximal exponent $\chi$ with a probability of one and the set of initial deviations for which we would obtain different Lyapunov exponents has zero measure \citep{skokos10}.

The maximal Lyapunov exponent $\chi$ corresponds to the most unstable direction in the phase space, however, exponents related to the complementary phase space directions can also be determined to reveal the whole Lyapunov spectrum. General properties of Lyapunov exponents imply that an autonomous Hamiltonian system of three degrees of freedom has four non-zero exponents out of which, however, only two are independent as they always appear in pairs of opposite signs but equal absolute values \citep{skokos10}. The computation of the whole spectrum of Lyapunov exponents can be performed using, e.g., the standard method developed by \citet{benettin80}. Although knowing the whole spectrum provides additional information about a given orbit, for the purpose of detecting chaotic dynamics it is sufficient enough to only compute the maximal exponent. For chaotic orbits, exponent $\chi$ attains a positive value\footnote{However, even in the integrable systems,  where no chaotic dynamics appears, {\em unstable periodic orbits} may be found, which also have positive Lyapunov exponents. Nevertheless, such orbits are rare as they form a set of zero measure \citep{contopoulos02}.}, while for regular trajectories it tends to zero as $\chi(\lambda)\propto\lambda^{-1}$. Therefore, in the logarithmic plot $\log\chi(\log\lambda)$, the regular trajectory appears as a linearly-decreasing function while the chaotic orbit, sooner or later, leaves the trend of linear decreasing and converges to the positive value. The Lyapunov time  $t_{\rm{L}}\equiv\chi^{-1}$ may thus be used to estimate the time after which the chaoticity of a given orbit manifests.\footnote{Lyapunov time is defined for the asymptotic value of $\chi$ as given by equation (\ref{mle}). In practical numerical applications we evaluate $t_{\rm{L}}$ of a chaotic orbit for such a value of $\lambda$ for which $\chi$ appears to attain its limit, i.e. it does not evolve anymore.} In particular, our operational criterion for {\em full} or {\em saturated} chaos will be the approximate relation  $t_{\rm{L}}\lesssim T_{\varphi}$, assuring that the trajectory exhibits chaotic properties on the time scale of its azimuthal period or even faster.

However, the convergence of $\chi$ may become very slow, and in some cases extremely long integration times are needed to reveal the chaotic nature of the trajectory. For example, some chaotic orbits get stuck in close vicinity of a regular orbit for a long time before they escape to the larger chaotic domain in the phase space \citep[so called ``stickiness effect'' --- see e.g.,][]{contopoulos10}. The computation of Lyapunov exponents may also be obstructed by numerical errors, especially when the long integration is required, the proper choice of the integrator becomes crucial \citep[e.g.,][]{seyrich12}. However, if one proceeds with caution, the method usually gives reliable results.

\begin{figure*}[htb!]
\centering
\includegraphics[scale=.8,clip]{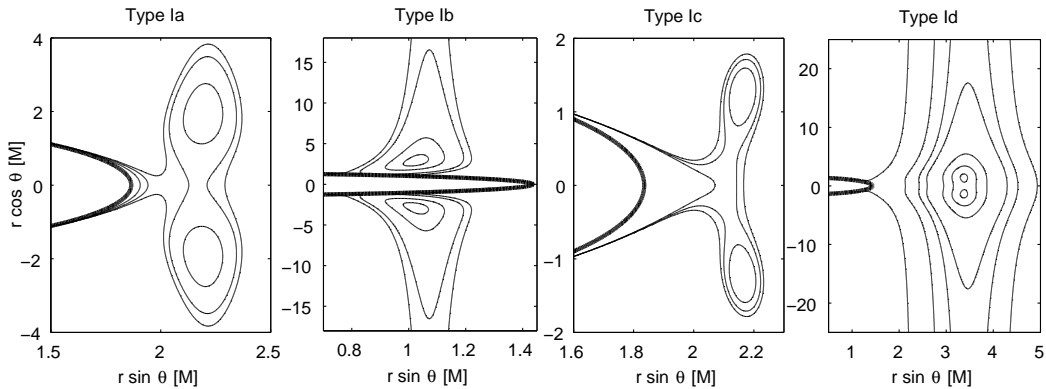}
\caption{Classification of possible topologies of off-equatorial potential lobes appearing above the event horizon (bold line) of the Kerr black hole immersed into the asymptotically-uniform magnetic field aligned with the rotation axis \citep{kopacek10b}.}
\label{wald_abcd}
\end{figure*}

In general, when performing the long-term integration of the Hamiltonian system, the method of choice would be a symplectic integrator \citep[e.g.,][]{yoshida93}. However, it is not possible to use a symplectic scheme for the simultaneous integration of the variational equations, and therefore we have to apply some non-symplectic method. Namely, we employ the Matlab integrator ODE87 which is an explicit 7th--8th order Runge--Kutta solver based on the Dorman--Prince formulas. The precision of this integrator with adaptive step-size is controlled  by setting the relative tolerance ($RelTol$), which specifies the highest allowed relative error in a single integration step. We were able to run our integrations with $RelTol=5\times10^{-16}$. During each integration we checked for the cumulative error (causing artificial excitation or damping of the system) by expressing the relative change of energy of the particle which maximally reached $\approx 10^{-5}$ for the longest runs.

\section{REGULAR AND CHAOTIC ORBITS}
\label{dynamics}
Common characteristics of previously investigated models are stationarity and axial symmetry. Such a system has two degrees of freedom in the four-dimensional spacetime. Considering the oblique magnetic field, however, we break the axisymmetry and the resulting system has three degrees of freedom, which has several important implications regarding the techniques we can apply for the analysis. First of all, we can no longer use the method of effective potential to localize the regions of bound orbits and, in particular, to find the stable, circular, off-equatorial orbits corresponding to the local minima of the potential (see Appendix~\ref{appa}). Besides that, the method of Poincar\'{e} surfaces of section,  which is commonly used to visualize the trajectories and distinguish between regular and chaotic dynamics in systems of two degrees of freedom, fails to deliver unambiguous results in this case. We will explore the behavior of the maximal Lyapunov characteristic exponent $\chi$ as a primary indicator of chaotic dynamics instead.

Inapplicability of the effective potential method leaves us without the usual tool for a systematical search for stable orbits throughout the configuration space and parameter space. In this place, we take advantage of our former study of axisymmetric versions of given systems \citep{kovar10,kopacek10} and use the particular, previously-investigated, confinements of particles in the system with aligned magnetic fields as the starting point of our current analysis. We will gradually incline the originally-aligned field and observe its impact upon the dynamics of particles. Such an approach has obvious limitations. We cannot set arbitrary inclination $B_x/B_z$ as we need to maintain the confinement of stable orbits whose presence at a given location non-trivially depends on its value. Actually, we shall fix all other parameters and increase the inclination of the field as long as we find the confinement of stable orbits. After reaching the critical inclination angle, the confinement disintegrates allowing the particles to escape to infinity or to fall onto the horizon. We are, however, only interested in the family of bound orbits. 

In the following we shall separately treat three distinct classes of orbits which we detected in the aligned field considering both charged and neutral black holes. Namely, in Section\ \ref{regoff} we shall explore the impact of inclination of the field upon the regular orbits in off-equatorial confinement. In Section\ \ref{chaoscross}, we consider chaotic orbits in the cross-equatorial lobe and in Section\ \ref{regeq}, the class of regular, equatorial trajectories is investigated. The distinction between equatorial and cross-equatorial orbits is based upon the different topology of effective potential defining the confinement. In the case of equatorial orbits, the underlying potential has a local minimum in the equatorial plane, while the cross-equatorial lobe is defined by the pair of symmetric off-equatorial minima separated by the saddle point residing in the equatorial plane. According to the results of our previous analysis, the latter configuration generally appears to host mainly chaotic orbits while the dynamics in the equatorial wells remains predominantly regular as long as the energy is raised only {\em slightly} above the level of the corresponding potential minimum. 

As we have already mentioned, we choose examples of both charged and neutral black holes. In general, the presence or absence of test charge $Q$ (generating Kerr--Newman test field which we superpose to the magnetic field --- see Equations (\ref{empot1})-(\ref{empot4})) does not seem to act as the factor systematically shifting the dynamics to become more or less chaotic. This is not surprising if we recall that the Kerr--Newman field alone does not represent a non-integrable perturbation for charged particles, as it admits the fourth integral of motion --- Carter's constant $\mathcal{L}$ \citep{carter68,mtw}. Although it evidently affects the dynamics of ionized particles (e.g., changes the location and shape of the particle confinements) it does not act as a trigger for chaos.

\subsection{Regular Off-Equatorial Orbits}
\label{regoff}
\begin{figure*}
\centering
\includegraphics[scale=0.5, clip]{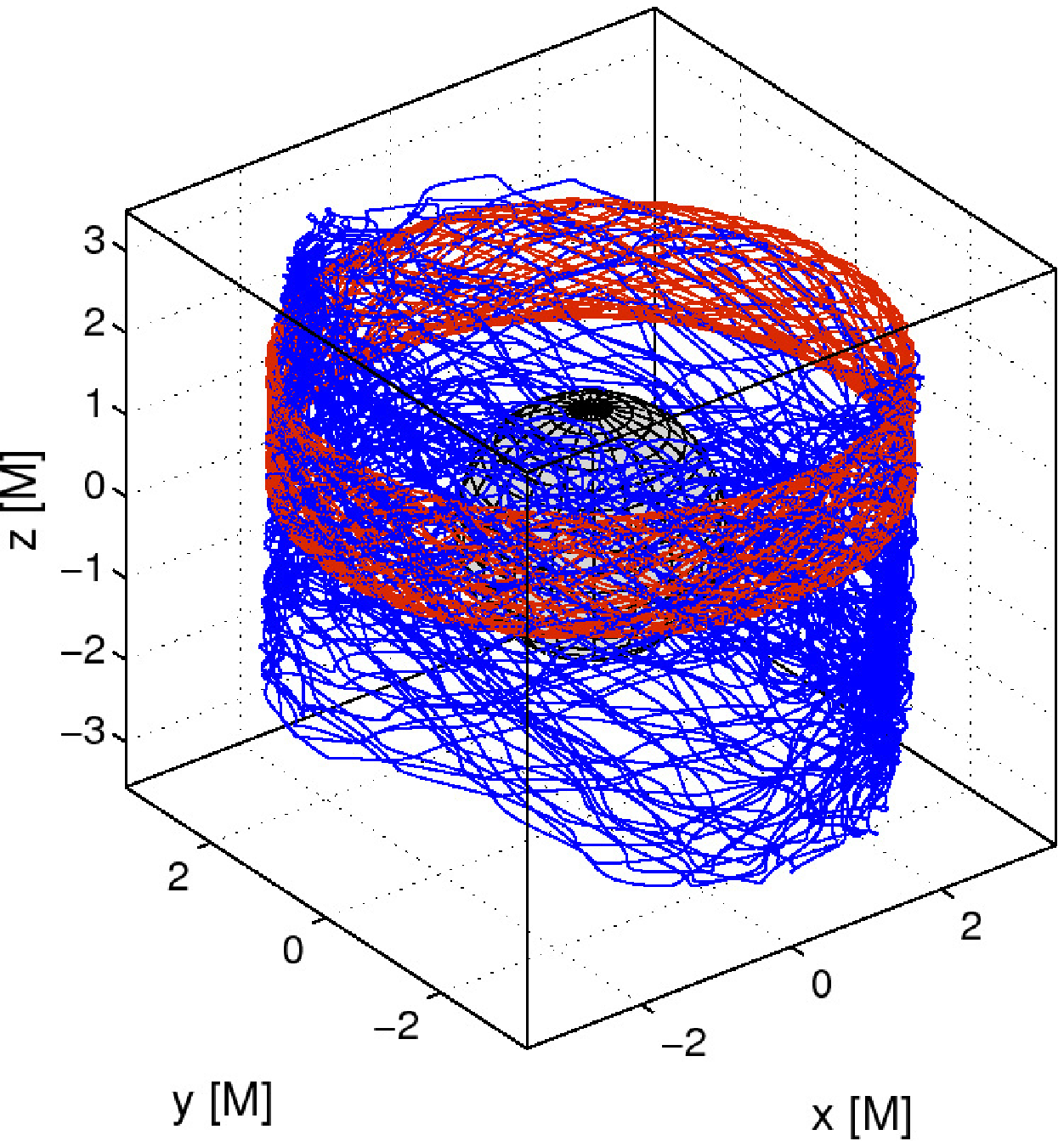}
\includegraphics[scale=0.46, clip]{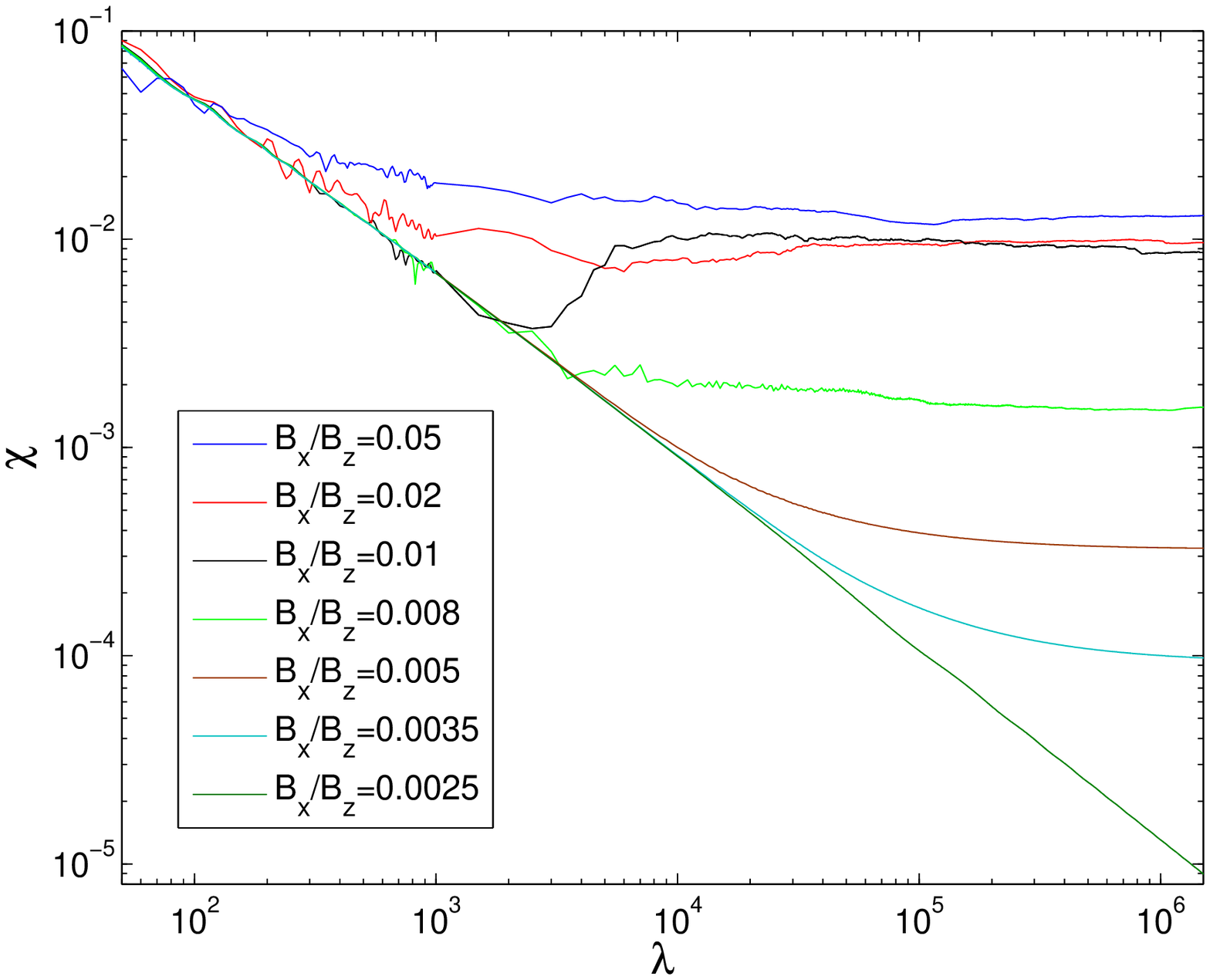}
\caption{Left panel shows the three-dimensional view of a regular off-equatorial orbit in the aligned field (red trajectory) which turns cross-equatorial and chaotic as the field inclines (blue orbit: $B_x/B_z=0.07$). Common parameters of the system are $a=0.9$, $E=1.58$, $qQ=1$, $qB_z=1$, and the initial condition is set as follows: $r(0)=3.68$, $\theta(0)=1.18$, $\varphi(0)=0$, $u^r(0)=0$ and $\pi_{\varphi}(0)=6$. Gray surface marks the horizon of the black hole. In the right panel we compare Lyapunov exponents of these trajectories which only differ in the inclination of the field. For a small value of the inclination angle the trajectory remains regular, however, for $B_x/B_z\gtrsim0.003$ the chaos sets on. Then the gradual growth of the largest Lyapunov exponent $\chi$ is observed as the inclination of the field increases. For $B_x/B_z\approx0.01$ the growth almost saturates. A logarithmic scale is used for the both axes.}
\label{off_reg_3D}
\end{figure*}
Considering the aligned magnetic field in \citet{kopacek10b}, we classified off-equatorial lobes according to the behavior of effective potential in the vicinity of its off-equatorial minima into four distinct classes Ia--Id (see \rff{wald_abcd}). Closed lobes define regions of possible confinement of particles, whereas the lobes extending down to the horizon allow particles to be accreted onto the black hole (types Ia--c). On the other hand, regions extending to large distance represent the case when particles can move far out (e.g., along the axis in type Id). Curves of constant energy suggest a different destiny of a particle on the off-equatorial circular orbit when its energy increases above the level of corresponding potential minimum. In type Ia the symmetric off-equatorial minima first merge via equatorial saddle point, allowing stable cross-equatorial motion. Increasing its energy further, the particle would reach another saddle point in the equatorial plane which allows it to fall freely onto the horizon. The class Ib does not permit any cross-equatorial motion --- off-equatorial confinements connect to the horizon directly when the energy increases sufficiently. Topology of type  Ic allows merging of off-equatorial lobes, however, at a merging point, these  are already opened toward the horizon and the motion in the cross-equatorial confinement is thus never stable. Type Id allows stable cross-equatorial motion. As the energy increases sufficiently the isopotential curve opens and particles may escape to infinity (unlike the case of Ia when they felt onto horizon instead).

Only two cases (namely types Ia and Id) exhibit the transition from the bound off-equatorial motion to the bound cross-equatorial motion when the energy is increased. Remaining types Ib and Ic do not allow closed cross-equatorial structures. In the following we analyze the impact of the inclination of the magnetic field on the originally regular off-equatorial orbit of type Id. In particular, to distinctly illustrate its influence we depart from the very same configuration as we used in \citet{kopacek10b}. Namely, we explored the orbit with parameters $L=6$, $a=0.9$, $qQ=1$, $qB_z=1$ and with the initial condition: $r(0)=3.68$, $\theta(0)=1.18$, $\varphi(0)=0$ and $u^r(0)=0$. For $E=1.58$ we observed regular off-equatorial motion while with $E=1.75$ the motion was cross-equatorial and chaotic. First we analyze the former case while the latter is inspected in Section\ \ref{chaoscross}. 

Gradually increasing the inclination $B_x/B_z$ and checking if the motion remains allowed and the orbit is still bound, we conclude that up to $B_x/B_z\approx0.05$ the allowed region remains closed, for values up to $ B_x/B_z\approx0.07$, it still exists as an open region allowing the particle to escape while even higher values forbid the motion for given parameters. In the left panel of \rff{off_reg_3D} we illustrate two examples of analyzed trajectories in the three-dimensional view. Most importantly, it shows that the inclination of the field extends the allowed region and converts the off-equatorial orbit into cross-equatorial in this case. Moreover, it suggests that the originally regular trajectory turns chaotic, which we confirm by means of the maximal Lyapunov exponent $\chi$ in  the right panel of \rff{off_reg_3D}.

For very small inclinations up to $B_x/B_z\approx0.003$, the trajectory remains regular and $\chi$ tends to zero limit  ($\log{\chi}$ decreases linearly as a function of $\log\lambda$). In particular, for the case of $B_x/B_z=0.0025$, we have checked this trend by integrating up to $\lambda=2\times10^7$. However, setting $B_x/B_z=0.0035$ the exponent $\chi$ attains a positive value which marks the onset of chaos. The asymptotic value of $\chi$ of the given orbit further rises as the inclination increases. For $B_x/B_z=0.01$, the trajectory becomes cross-equatorial and $\chi$ almost saturates as it does not significantly rise when the field is further inclined. The trajectory is now {\em fully chaotic} as its Lyapunov time roughly equals the orbital period $t_{\rm{L}}\approx T_{\varphi}\approx100$ (numerically averaged along the actual orbit --- see Appendix~\ref{appb}). Values $B_x/B_z>0.05$ correspond to the opened allowed region and the particle typically escapes before $\lambda=10^5$. 

The analyzed example reveals the typical behavior of the originally regular off-equatorial orbits perturbed by inclining the magnetic field. Most importantly, it shows that the inclination of the field induces continuous transition to chaos as measured by means of the maximal Lyapunov exponent $\chi$. 

\subsection{Chaotic Cross-equatorial Orbits}
\label{chaoscross}
\begin{figure*}
\centering
\includegraphics[scale=0.55, clip]{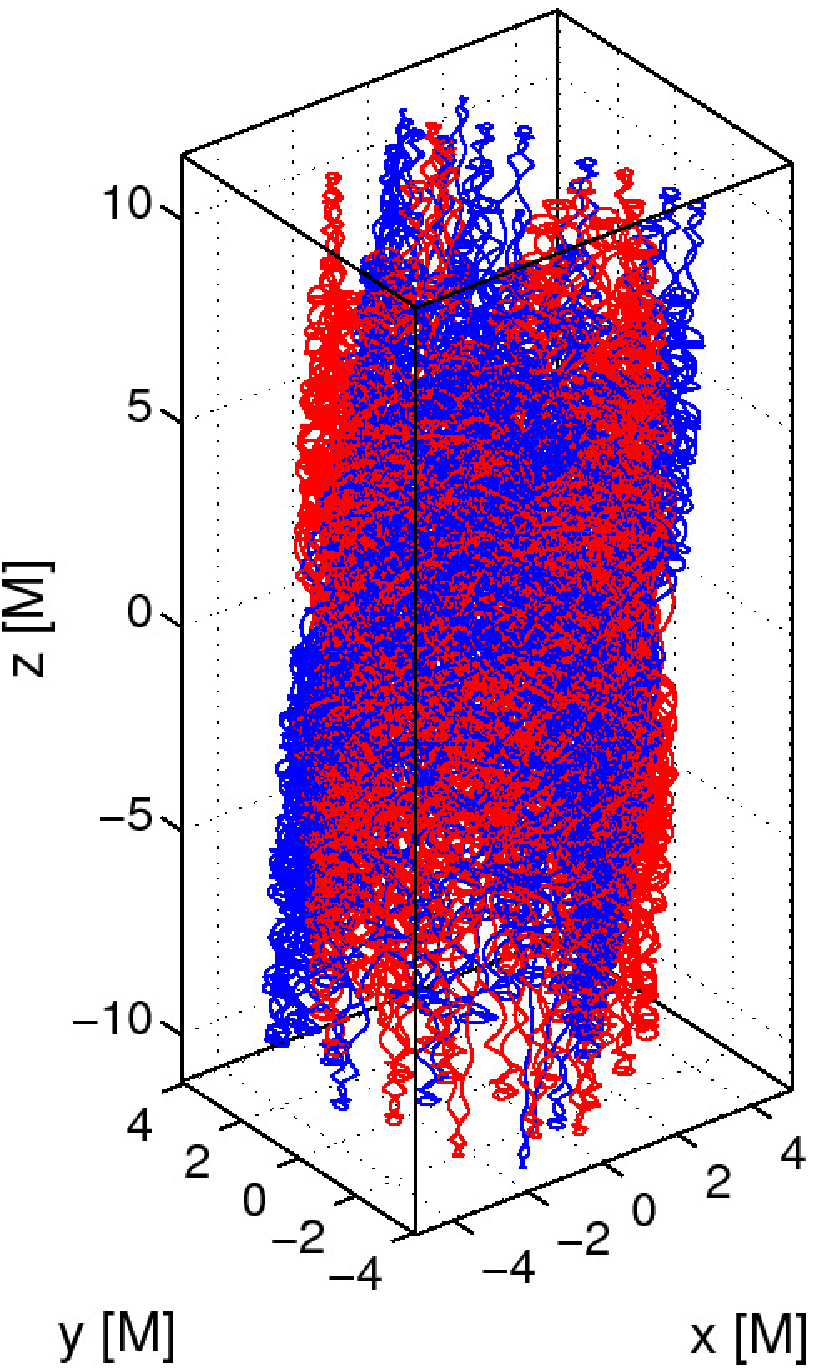}
\includegraphics[scale=0.54, clip]{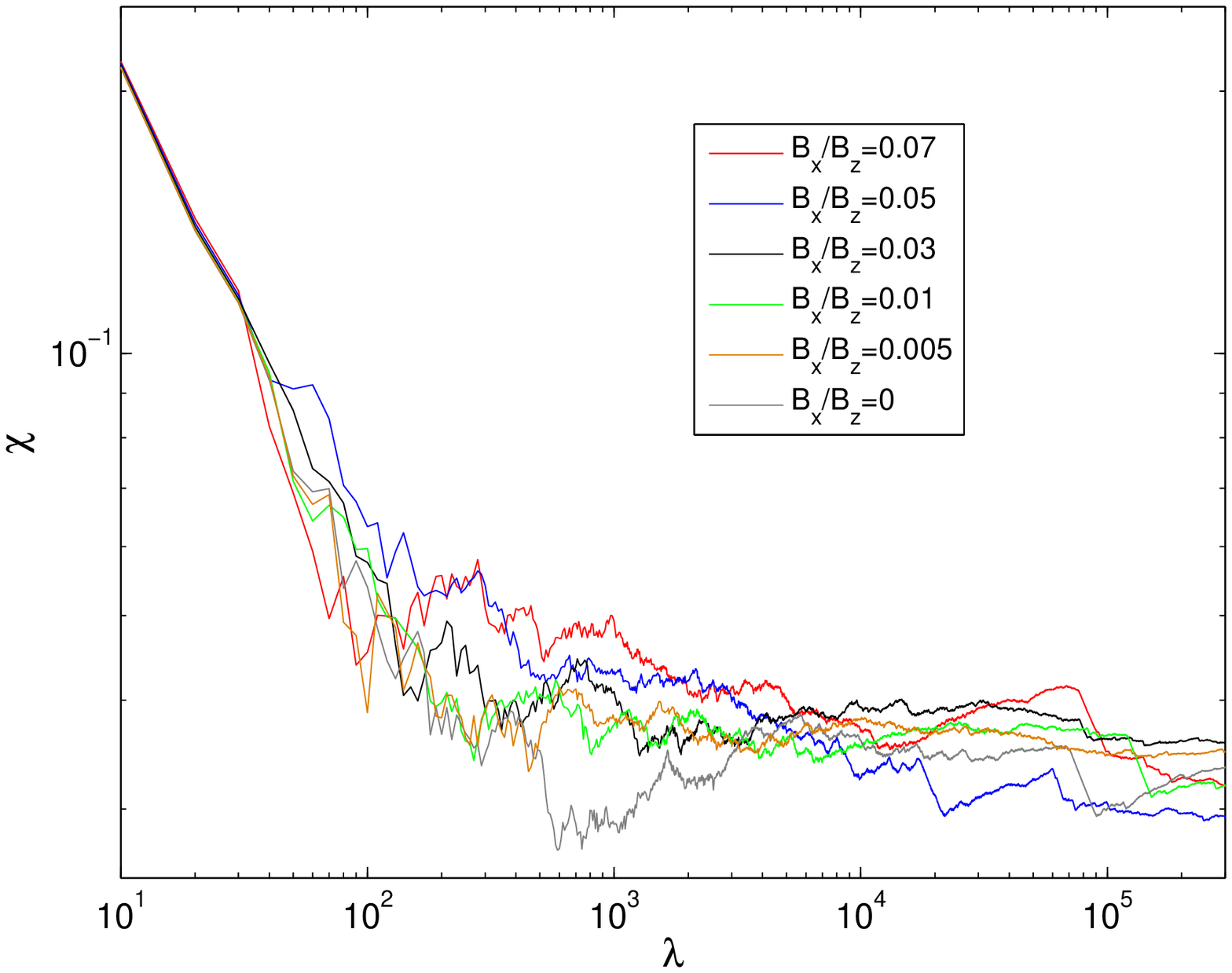}
\caption{In the left panel two chaotic cross-equatorial orbits with different inclinations of the background magnetic field are shown. The red trajectory is followed by the particle in the aligned field ($B_x/B_z=0$), while the blue one is driven by the considerably-inclined field ($B_x/B_z=0.15$). Energy is set as $E=1.75$ while other parameters are kept at values specified in \rff{off_reg_3D}. Besides the inclination of the whole trajectory, the orbits under comparison do not exhibit any significant qualitative difference. Fully chaotic orbit appears dynamically indifferent to the inclination of the field which is confirmed in terms of corresponding Lyapunov exponents of these trajectories in the right panel. We observe that $\chi$ does not considerably change when the inclination increases. Unlike the previous case where the regular trajectory was perturbed by inclining the field, here the values of $\chi$ do not produce an ordered sequence (higher inclination does not mean higher $\chi$ in general) and remain almost indifferent to the field's inclination. Chaoticity of the fully chaotic trajectory ($t_{\rm L}\approx T_{\varphi}$) cannot be further increased by an additional perturbation.}
\label{ekv_chaos_3D}
\end{figure*}
Here we inspect the impact of the inclination on the cross-equatorial orbit which is already chaotic in the aligned field. We choose to analyze the example analogical to the previous case of the off-equatorial orbit. Namely, we start with the trajectory which has a higher energy $E=1.75$ while all other parameters and initial conditions remain fixed at values specified above. In the left panel of \rff{ekv_chaos_3D}, we compare the three-dimensional views of two typical chaotic orbits: the original aligned trajectory (red) and an inclined trajectory with $B_x/B_z=0.15$ (blue). Higher inclinations are not permitted for given setup. Inclined field lines are traced by the trajectory which clearly shows the direction of the field. However, besides the inclination of the trajectory there is no apparent qualitative difference between compared orbits in \rff{ekv_chaos_3D}.

Quantitative comparison in terms of the Lyapunov exponent $\chi$ confirms that, in this case, the inclination of the magnetic field does not significantly change the dynamics (right panel of \rff{ekv_chaos_3D}). We observe that unlike the previous case of off-equatorial regular orbits, here the asymptotic values of $\chi$ are not stratified with respect to the inclination angle. A higher inclination angle does not necessarily correspond to a higher maximal Lyapunov exponent $\chi$.  Changing the inclination between $B_x/B_z=0$ and $B_x/B_z=0.07$ results in comparable asymptotic values of $\chi$. We conclude that once the system is fully chaotic, the additional perturbation consisting in the inclination of the field does not increase the chaoticity of the system.

\subsection{Regular Equatorial Orbits}
\label{regeq}
We demonstrate the influence of an inclined magnetic field on the originally regular orbit in the equatorial potential well. Here we choose the orbit around the black hole without an additional electric charge as a representative example.

Parameters of the trajectory are $a=0.9$, $E=1.24$, $qQ=0$, $qB_z=1$, and the initial condition is set as follows: $r(0)=3$, $\theta(0)=\pi/2$, $\varphi(0)=0$, $u^r(0)=0$, and $\pi_{\varphi}(0)=5$. In the given setup the trajectory remains bound for inclinations up to $B_x/B_z\approx0.1$ and the motion is allowed for $B_x/B_z\lesssim0.15$. The three-dimensional view of the aligned trajectory (red) and two orbits in the oblique field with different inclinations (green: $B_x/B_z=0.05$, blue: $B_x/B_z=0.1$) are compared in the left panel of \rff{ekv_reg_3D}. For the aligned field, this orbit exhibits regular dynamics and the corresponding trajectory draws an axisymmetric pattern. The oblique field gradually {\em inclines} the trajectory as a whole and introduces apparent signs of disorder whose symptoms increase with the increasing inclination.

The comparison of Lyapunov exponents $\chi$ corresponding to orbits with different inclinations is given in the right panel of \rff{ekv_reg_3D}. In this case, only the aligned trajectory remains truly regular with $\log\chi$ falling linearly to zero as a function of $\log\lambda$. Even for a very small inclination of $B_x/B_z=0.0005$, the trajectory eventually leaves the trend of linear decrease of $\log\chi$ and approaches a positive value in the limit, which is a hallmark of the chaotic dynamics. With the growing inclination, the limiting value of $\chi$ gradually grows. More inclined field triggers {\em stronger chaos} as measured by the exponent $\chi$. We notice that in this case, the convergence of $\chi$ is remarkably smooth for all allowed inclinations which we observed in the case of off-equatorial orbits (\rff{off_reg_3D}) only for smaller inclinations which were inducing non-saturated chaos.

The given example of the originally regular equatorial orbit affected by inclining the magnetic field, illustrated high sensitivity of regular dynamics upon the axial symmetry of the system. In this case, breaking the symmetry inevitably brings chaotic features to the system regardless the value of the inclination.

\begin{figure*}
\centering
\includegraphics[scale=0.46, clip]{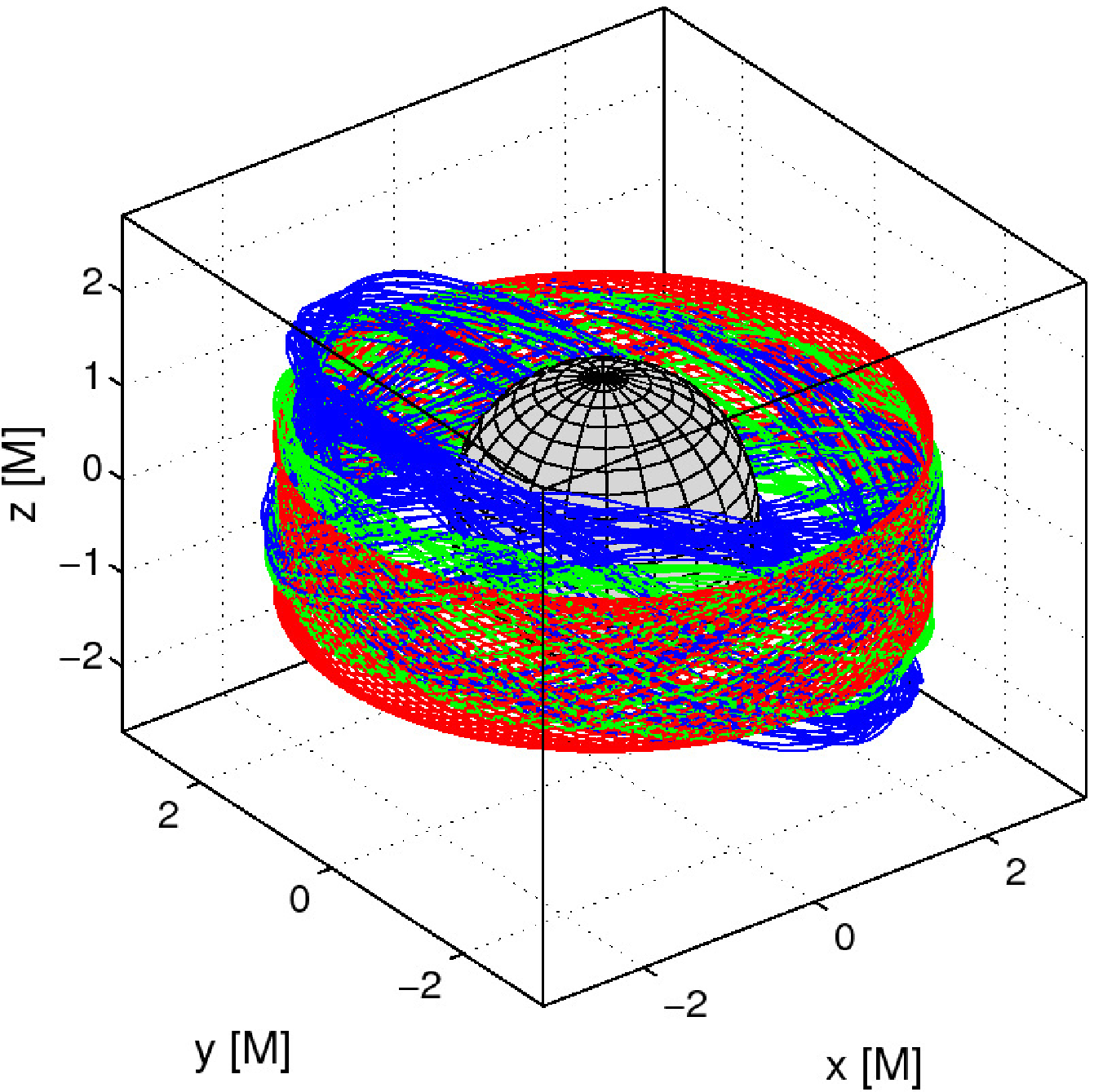}
\includegraphics[scale=0.46, clip]{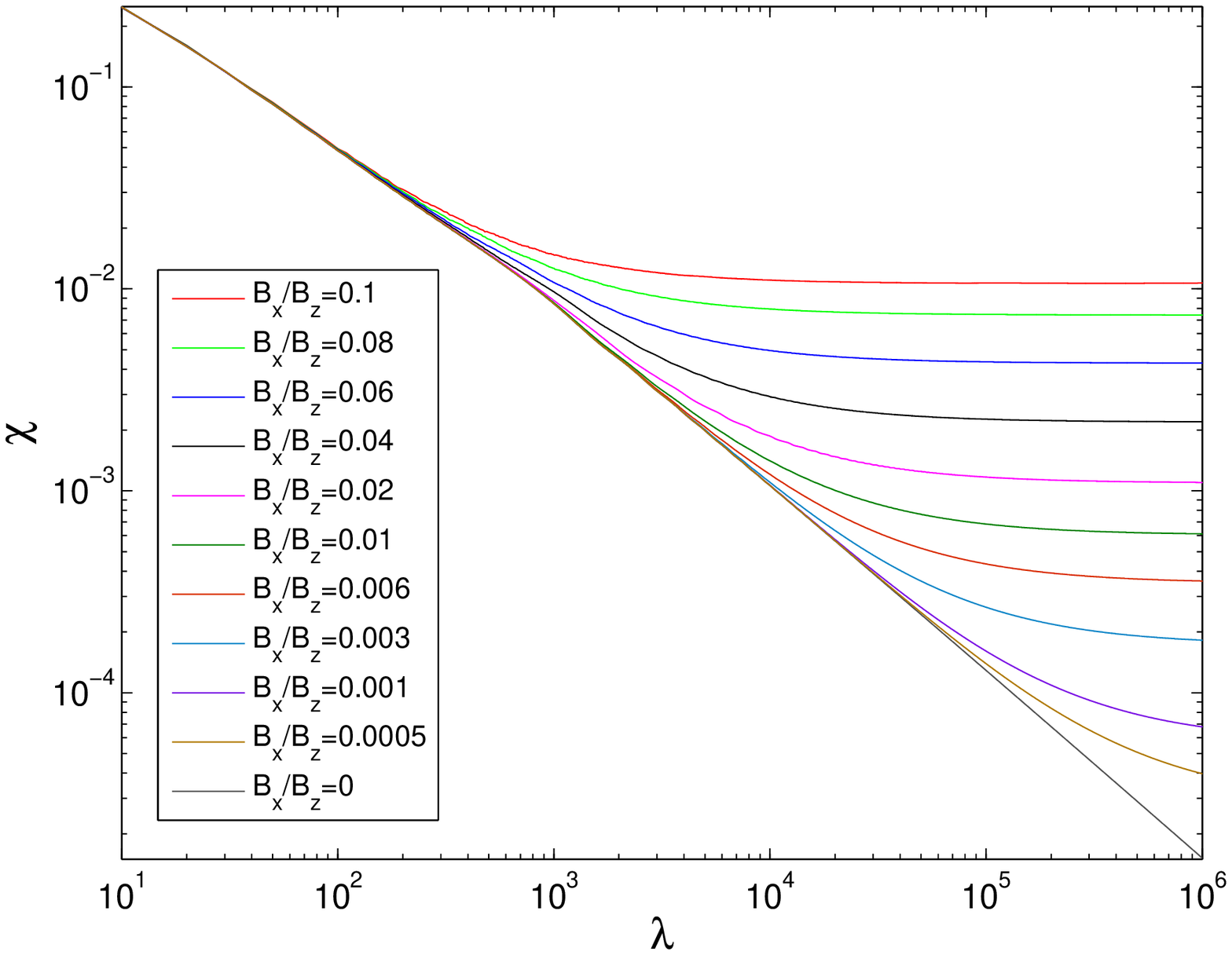}
\caption{Three-dimensional view of three equatorial orbits differing in the inclination of the background magnetic field is presented in the left panel. The red orbit is regular and it is followed by the particle in the aligned field ($B_x/B_z=0$), while for the inclined fields we detect chaotic dynamics (green: $B_x/B_z=0.05$, blue: $B_x/B_z=0.1$). Parameters of the system are $a=0.9$, $E=1.24$, $qQ=0$, $qB_z=1$ and the initial condition is set as follows: $r(0)=3$, $\theta(0)=\pi/2$, $\varphi(0)=0$, $u^r(0)=0$ and $\pi_{\varphi}(0)=5$. Gray surface marks the horizon of the black hole. In the right panel we compare the behavior of the maximal Lyapunov exponents corresponding to these trajectories. Even for very small inclination angles the orbit turns chaotic. Gradual growth of the largest Lyapunov exponent $\chi$ is observed as the inclination of the field increases. A logarithmic scale is used for both axes.}
\label{ekv_reg_3D}
\end{figure*}

In \rff{lya_vs_uhel} we compare asymptotic values of $\chi$ as a function of the inclination of the field for all three types of orbits analyzed above and illustrated separately in Figs.~\ref{off_reg_3D}--\ref{ekv_reg_3D}. The originally off-equatorial, regular orbit (red curve) shows a fast onset of chaos as the inclination increases and cross-equatorial motion becomes possible. However, the growth of $\chi$ quickly saturates as the system finishes the transition to the chaotic regime (at $B_x/B_z\approx0.01$) and, beyond this point, the value of $\chi$ does not grow considerably with the increasing inclination of the field. Departing from the cross-equatorial orbit which is already fully chaotic in the aligned field $B_x/B_z=0$, we observe no systematic reaction of $\chi$ when the field gradually inclines (blue curve). The system in fully chaotic mode does not respond to the additional perturbation by increasing its chaoticity. The originally regular equatorial orbit (green plot) undergoes a gradual growth of chaoticity as the field inclines. Nevertheless, the value of $\chi$ does not saturate within the range of allowed inclinations ($B_x/B_z\lesssim0.15$).

\begin{figure}[htb!]
\centering
\includegraphics[scale=0.46, clip]{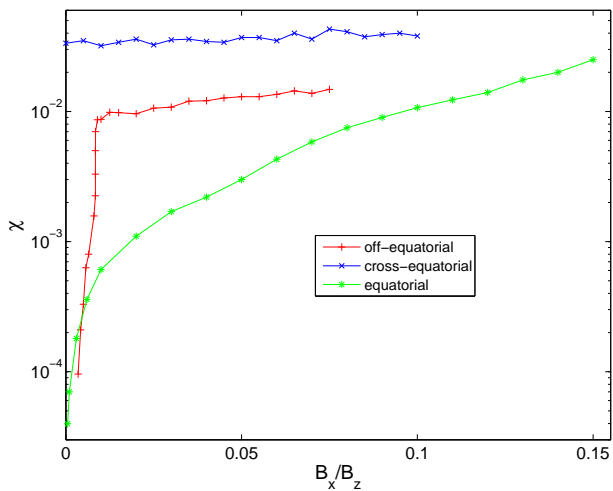}
\caption{Asymptotic value of the maximal Lyapunov exponent $\chi$ (in the logarithmic scale) as a function of the inclination of the field $B_x/B_z$ (tangent of the inclination angle) for three representative cases which were discussed in Figs.~\ref{off_reg_3D}--\ref{ekv_reg_3D}. The red curve shows the behavior of the exponent $\chi$ of the originally regular off-equatorial orbit (parameters specified in \rff{off_reg_3D}). The response of the cross-equatorial chaotic orbit is plotted by the blue curve (orbit's parameters  in \rff{ekv_chaos_3D}), and the originally regular equatorial orbit from \rff{ekv_reg_3D} is represented by the green line. We note that the inclinations we considered in the analysis generally correspond to small inclination angles of the field given as $\arctan(B_x/B_z)$, e.g., the inclination of $B_x/B_z=0.1$ corresponds to the angle of $\approx6^{\circ}$.}
\label{lya_vs_uhel}
\end{figure}

\section{DISCUSSION}
\label{discussion}
The scenario investigated within this paper is idealized and its direct applicability for interpretation of observed astrophysical phenomena is limited. However, the significance of our model lies in its ability to demonstrate the effect of combined magnetic and strong gravitational fields onto the dynamics of charged matter in a very clear form since the matter is treated as collisionless and thus its motion is governed solely by external fields. Under such circumstances we raised the fundamental question whether a given combination of fields typically leads to chaotic or regular dynamics of orbiting particles.  

In the context of this analysis another natural question arises ---  whether it is possible to observationally distinguish imprints of ordered or chaotic dynamics occurring in the investigated system. In general, the link between chaos on a microscopic scale and the macroscopic behavior of matter has not yet been fully understood and it represents an open problem of statistical mechanics. While, in some systems, the relation of transport coefficients (e.g., viscosity, thermal, or electric conductivity, etc.) to Lyapunov exponents is observed, there are also non-chaotic systems that show transport phenomena and thus suggest that chaos might not be necessary for the robust statistical behaviors  \citep[see e.g.,][and references therein]{castiglione08, cencini08, cencini10}. 

Nevertheless, we are in a slightly different situation as we do not build a statistical ensemble of particles to derive its macroscopic properties in the framework of kinetic theory. As we rather consider a set of non-interacting particles, can we expect them to give characteristic imprints of their dynamic regime on the generated radiation? We believe that the answer is, at least in principle, positive. Frequency analysis of bound trajectories shows a clear difference between order and chaos \citep[][]{contopoulos02, lukes10}. While regular trajectories have discrete power spectra with prominent lines at their fundamental frequencies, the chaotic ones instead show continuous spectra. In the case of charged particles which radiate when accelerated, such distinction in the spectra of dynamic frequencies inevitably translates into a frequency composition of generated electromagnetic signals. Determining the actual form of resulting electromagnetic spectra would, however, require a detailed analysis, which is beyond the scope of this paper. 

We also note that the general problem of detecting chaos and quasiperiodicity in experimental data with noise has been  studied thoroughly \citep[see e.g.,][]{sprott03,nayfeh95} and a number of relevant methods have been developed over the last decades. Not listing all standard techniques, we only mention a more recent method of recurrence analysis \citep{marwan}, which appears very useful for this task.  

The crucial question remains: whether the electromagnetic signal generated by our system is strong enough to deliver characteristic imprints of order and chaos to the observer. In Appendix~\ref{appb} we thus calculate the power $P$ of synchrotron radiation generated by charged particles orbiting in the confinements discussed above. The total power of this radiation as given by Equation (\ref{powertot}) is negligible. The obvious way to substantially enhance the power output is to accelerate the particles to ultrarelativistic velocities since $P$ scales with Lorentz factor $\gamma$ as $P\propto \gamma^4$. In our model, however, the velocity (\ref{linspeed}) is not a free parameter and it seems that ultrarelativistic velocities are beyond its scope. Estimating the bremsstrahlung resulting from Coulomb collisions (free--free radiation) by Equation (\ref{brem}), we find this radiative process to be substantially more efficient in given circumstances. However, unlike synchrotron radiation, we do not expect the bremsstrahlung to encode imprints of dynamic regime of colliding particles in its spectrum.

We have shown that a charged matter in densities consistent with our collisionless description does not emit a significant amount of synchrotron radiation in which we could distinguish imprints of regular or chaotic dynamic regime of emitting particles. The mechanism of inverse Compton scattering of thermal photons emitted from the accretion disk on electrons trapped in the confinements discussed within this paper may be more effective in delivering observational imprints in detectable signals (ongoing work). The main motivation of our current analysis was to study the chaoticity in the  dynamics of charged matter in accreting black hole systems as a fundamental theoretical aspect of its physics.
\newpage

\section{CONCLUSIONS}
\label{conclusions}
Regularity is a remarkable property of motion around black holes. We addressed the question of transition from regular to chaotic dynamics by including the effect of magnetic field, and we studied the importance of misalignment between the direction of the magnetic field and the spin of the black hole.

In this paper we investigated some aspects of dynamics in the test particle model of the black hole magnetosphere. In particular, we considered a system consisting of a rotating black hole embedded in the large-scale ordered magnetic field. In our previous studies \citep{kovar10,kopacek10} we found that, in the special case of an aligned field (i.e., the magnetic field was symmetric with respect to the rotation axis), the motion of matter in off-equatorial confinements was typically regular. The chaotic dynamics usually appeared for cross-equatorial orbits. Considering the equatorial potential wells instead, we observed the dominance of regular dynamics for energies in a certain limited range, followed by a continuous transition to the mostly-chaotic regime  when the energy was gradually increased above this range. 

Here we generalized our model by considering oblique magnetic fields. In particular, we investigated the impact of the inclination upon the dynamics of charged matter orbiting around a black hole in three distinct types of confinements, namely in off-equatorial, cross-equatorial, and equatorial lobes. Using the maximal Lyapunov exponent $\chi$ as a basic indicator of chaotic dynamics, we found that regular orbits are, in general, highly sensitive to the perfect alignment of the field. The inclination of $B_x/B_z\approx 0.01$ ($\approx30'$) was typically sufficient to perturb the regular dynamics and introduce prominent chaotic features which we illustrated in several representative examples. In particular, we found that in the case of originally regular, off-equatorial, motion the system remains regular only for very small inclinations (up to $B_x/B_z\approx 0.003$ in analyzed example) and with increasing inclination it quickly shifts to the chaotic regime. Considering the class of regular equatorial orbits instead, we observed that these did not oppose the perturbation at all as they became chaotic for inclinations as low as $B_x/B_z\approx 0.0005$. Also in this case the system undergoes gradual transition to full chaos as the inclination increases. On the other hand, if the original trajectory is already fully chaotic in the aligned configuration, the further perturbation to the system by inclining the field does not significantly affect the dynamics. The trajectory as a whole inclines along the field (see \rff{ekv_chaos_3D}), however, the chaoticity as measured by the Lyapunov exponent $\chi$ does not increase further.

We conclude that within the given model the stability of the regular motion of charged particles depends critically on the the perfect alignment of the large-scale magnetic field with the rotation axis of the black hole. Once the field is slightly inclined and the symmetry of the system breaks, the chaotic regime dominates the dynamics.   

\acknowledgements
O.K. is grateful to obtain support from the Czech Science Foundation (GA \v{C}R ref.\ 202/09/0772) and acknowledges the postdoctoral program of the Czech Academy of Sciences. V.K. acknowledges the Czech Science Foundation (13-000703).

\appendix
\section{EFFECTIVE POTENTIAL}
\label{appa}
In the classical mechanics, the notion of the effective potential is usually introduced when dealing with the central force problem which concerns the motion of the particle of negligible mass ({\em test particle}) which is attracted or repelled by a static massive center. The magnitude of this force depends solely on the distance from the center. A prominent instance of such a setup in which the central force is represented by the gravitational attraction is a Kepler problem (and a related gravitational two-body problem which can be reformulated as two one-body problems, one trivial and the other of the Kepler type). Effective potential $V_{\rm eff}$ is then given as a sum of the gravitational potential and the centrifugal term related to the angular momentum $L$ of the orbiting body. Analysis of the effective potential provides valuable overall information about the dynamics without needing the actual integration of particular trajectories. Its significance arises from the defining relation $\epsilon-V_{\rm eff}=T$ where $\epsilon$ is the classical total energy of the particle and $T$ represents its kinetic part which comprises only the radial term $m\dot{r}^2/2$ in this case since $\dot\varphi$ is expressed in terms of $L$. Thus, the effective potential (as a function of $r$ and $L$) expresses the energy of the particle at which the radial turning points occur (apocenter and pericenter in the case of bound orbits). It defines the boundary (in the extended configuration space) of the regions of the allowed motion. In particular, it allows us to locate stable, circular orbits which correspond to the minima of the effective potential.

In general relativity, however, we have no clear distinction between the kinetic and potential energy. Nevertheless, in many cases we may still derive function analogous to the classical effective potential which allows us to explore the dynamics of test particles and, in particular, to locate regions of stable orbits.  In the general case of the charged test particle of the rest mass $m$ and charge $q$ in the spacetime with metric $g^{\mu\nu}$ and electromagnetic field $A_{\mu}$ we depart from the super-Hamiltonian $\mathcal{H}$ whose conserved value is given by the normalization of the four-momentum:
\begin{equation}
\label{fourmomentum} 2\mathcal{H}=g^{\mu\nu}p_{\mu}p_{\nu}=g^{\mu\nu}(\pi_{\mu}-qA_{\mu})(\pi_{\nu}-qA_{\nu})=-m^2.
\end{equation}

In the special case of the stationary and axisymmetric background of a Kerr or Kerr--Newman black hole with an additional, stationary, electromagnetic test field obeying the same symmetry in which $\pi_{\varphi}=L$ and $\pi_t=-E$ are constants of motion (system therefore has two degrees of freedom) and neither $g_{\mu\nu}$ nor $A_{\mu}$  depend on Boyer--Lindquist coordinates $t$ and $\varphi$, we obtain by straightforward manipulations 
\begin{equation}
\label{eff}
\Sigma\left(\frac{(p^{r})^2}{\Delta}+(p^{\theta})^2\right)=\alpha{}E^2+\beta{}E+\gamma,
\end{equation}
where
\begin{eqnarray}
\label{coeff}
{\nonumber}\alpha&=&-g^{tt}\\
\beta&=&2\left[g^{t\varphi{}}(L-qA_{\varphi})-g^{tt}qA_{t}\right]\\
{\nonumber}\gamma&=&-g^{\varphi\varphi}(L-qA_{\varphi})^2-g^{tt}q^2A_{t}^2+2g^{t\varphi{}}qA_{t}(L-qA_{\varphi})-m^2.
\end{eqnarray}
Since both coefficients $\Sigma$ and $\Delta$ are positive above the outer horizon $r_+$ to which region we restrict our study, the zero point of the left hand side of (\ref{eff}) occurs at the simultaneous turning point of motion in both the radial and latitudinal direction, and defines the boundary of allowed motion. A function which specifies the value of energy corresponding to the turning point can be regarded as a generalization of the classical effective potential $V_{\rm eff}$. We can therefore express the two-dimensional effective potential $V_{\rm eff}(r,\theta)$ as follows:
\begin{equation}
\label{effpot}
V_{\rm eff}=\frac{-\beta+\sqrt{\beta^2-4\alpha\gamma}}{2\alpha},
\end{equation}
where the positive square root has to be chosen to correspond with the future-pointing four-momentum \citep[][p. 909]{mtw}. Since $\alpha>0$ above the horizon the motion is allowed just if $E\geq V_{\rm eff}$.

A method of effective potential has been applied to locate confinements (both equatorial and off-equatorial) of charged matter in several stationary and axisymmetric models in our previous works \citep{kovar13,kopacek10, kovar10, kovar08}. Potential $V_{\rm eff}$ was investigated as a function of two configuration variables $r$ and $\theta$, angular momentum $L$, and particular parameters of the given system. 

In the context of present paper, our question is whether we could also apply this method for the system of three degrees of freedom in which the axial symmetry is broken and $A_{\mu}=A_{\mu}(r,\theta,\varphi)$ but $g_{\mu\nu}=g_{\mu\nu}(r,\theta)$. In this case, the trajectory manifold spans five out of eight dimensions of the phase space due to the stationarity and autonomy of the system. Effective potential reduces the number of dimensions by imposing the constraint of type $(p^{\mu})^2=0$ which locates the turning point in $\mu$-direction. In this case, we seek the simultaneous turning point in all three directions $r$, $\theta$, and $\varphi$ which would result in two-dimensional submanifold. For a fixed value of $\varphi$ we should therefore obtain one-dimensional isopotential curves specifying the allowed region in a given meridional plane described by coordinates $r, \theta$ as we previously did in the case of axisymmetric systems. Indeed, we can derive the expression formally analogous to (\ref{eff}):

\begin{equation}
\label{eff2}
\Sigma\left(\frac{(p^{r})^2}{\Delta}+(p^{\theta})^2\right)+g_{\varphi\varphi}(p^{\varphi})^2=\alpha^{\star}E^2+\beta^{\star}{}E+\gamma^{\star},
\end{equation}
where the coefficients are now given as
\begin{eqnarray}
\label{coeff2}
{\nonumber}\alpha^{\star}&=&-g^{tt}\left(1+g^{t\varphi}g_{t\varphi}\right)\\
\beta^{\star}&=&2\left[g_{t\varphi}(g^{t\varphi})^2(\pi_{\varphi}-qA_{\varphi})-g^{tt}qA_{t}(1+g^{t\varphi}g_{t\varphi})\right]\\
{\nonumber}\gamma^{\star}&=&-g^{\varphi\varphi}g^{t\varphi}g_{t\varphi}(\pi_{\varphi}-qA_{\varphi})^2-g^{tt}q^2A_{t}^2(1+g^{t\varphi}g_{t\varphi})+2(g^{t\varphi{}})^2g_{t\varphi}qA_{t}(\pi_{\varphi}-qA_{\varphi})-m^2.
\end{eqnarray}

The left-hand side of (\ref{eff2}) has the proper form necessary for expressing the effective potential ($g_{\varphi\varphi}$ is positive). Nevertheless, the coefficients $\beta^{\star}$ and $\gamma^{\star}$ depend on the azimuthal component of canonical momentum $\pi_{\varphi}$ which used to be the integral of motion $L$ in the axisymmetric system, however, here it changes along the trajectory. The evolution of $\pi_\varphi$ is not known a priori and one has to integrate the equations of motion of a given particle to reveal it. Therefore, it is not possible to express the effective potential from the above equation as a function of $r$, $\theta$, and $\varphi$ coordinates (and parameters of the metric and electromagnetic field). We conclude that the given technique leads to the derivation of the effective potential only in the case of aligned magnetic fields. For oblique fields, the method fails to provide the potential since the simultaneous turning points and the boundaries of allowed regions are actually not captured by resulting formula (\ref{eff2}).  

\clearpage
\section{RADIATION POWER}
\label{appb}
It has been proposed that a distinction between chaos and regular motion in a system of radiating particles can be revealed in the power spectra of the resulting signal. Regular trajectories will contribute to localized features in the power-density spectrum (PDS), which disappear when chaos prevails. To assess the detectability, we first need to estimate the strength of the outgoing signal from the system.

Here we estimate the power of synchrotron radiation generated by charged particles orbiting in confinements discussed in this paper. To this end we make approximations whose appropriateness will be discussed. 

First we pick one individual trajectory ({\em typical orbit}) whose properties will characterize the whole ensemble of particles in a given confinement. We choose an equatorial trajectory with parameters $a=0.9$, $E=1.24$, $qQ=0$, $qB_z=1$, $qB_x=0$, and initial condition $r(0)=3$, $\theta(0)=\pi/2$, $\varphi(0)=0$, $u^r(0)=0$ and $\pi_{\varphi}(0)=5$ (\rff{ekv_reg_3D}). For this trajectory, we determine linear velocity $v^{(i)}$ with respect to the locally non-rotating frame \citep{bardeen72} with the tetrad basis $e^{(i)}_{\mu}$ as follows
\begin{equation}
\label{linspeed}
 v^{(i)}=\frac{u^{(i)}}{u^{(t)}}=\frac{e^{(i)}_{\mu}u^{\mu}}{e^{(t)}_{\mu}u^{\mu}}.
\end{equation}
Calculating $v^{(i)}$ over a sufficiently-long integration period ($\approx 100$ revolutions around the center) we find that the mean of the azimuthal component, which dominates the motion, reads $v_{\perp}\equiv \left|\left< v^{(\varphi)}\right>\right|\approx 0.1c$. Since we only consider small inclinations of the magnetic field, the azimuthal component $v^{(\varphi)}$ approximately equals the component perpendicular to the field $v_{\perp}$ also in the case of oblique fields. The corresponding Lorentz factor reads $\gamma\approx1.005\approx 1$ which justifies the cyclotron regime.

As a next step, we specify the charge of the particles under consideration as well as the strength of the magnetic field,
\begin{equation}
 \label{magfield}
B_{\rm SI}=\frac{qB_zc_{\rm SI}}{q_{\rm SI}\left(\frac{M}{M_{\odot}}\right)1472\;\rm{m}},
\end{equation}
where the quantities without subscript ${\rm SI}$ are dimensionless (expressed in geometrized units and scaled by mass of the black hole $M$) and the length $1472\;\rm{m}$ reflects the solar mass in geometrized units, $M_{\odot}=1472\:\rm{m}$.
Inserting the value $qB_z=1$ and fixing the specific charge $q_{\rm SI}$ to electron, i.e. $q_{\rm SI}=1.76\times10^{11}\; \rm{C}\,\rm{kg}^{-1}$, we find that for the black hole of $M=10\;M_{\odot}$, the corresponding magnetic field reads $B_{\rm SI}=1.16\times10^{-7}\;\rm{T}$. Such a value is consistent with non-thermal filaments in the Galactic Center \citep{ferr10,larosa04}.

The Larmor radius of the particles' gyration $r_{\rm L}$ is then given as follows
\begin{equation}
 \label{gyro}
\left(r_{\rm L}\right)_{\rm SI}=\frac{1472\;\rm{m}}{qB_z}\left(\frac{v_{\perp}}{c}\right)\left(\frac{M}{M_{\odot}}\right),
\end{equation}
which reads $\left(r_{\rm L}\right)_{\rm SI}=1472\;{\rm m}$ for electron with $v_{\perp}=0.1c$ and $M=10\;M_{\odot}$. However, we stress that, in our case, it is not the actual radius of the orbit which is determined not only by the magnetic field, but also by the gravitational pull of the black hole. Indeed, the radius of our typical orbit roughly equals $r(0)=3$ which corresponds to $r_{\rm SI}=3(M/M_{\odot})1472\;\rm{m}\approx44\;{\rm km}$. The azimuthal period of a typical orbit (computed as a mean of $\approx 100$ revolutions) reads $\left(T_{\varphi}\right)_{\rm SI}\approx0.26\left(M/M_{\odot}\right)\;{\rm ms}=2.6\;\rm{ms}$ while the latitudinal period is $\left(T_{\theta}\right)_{\rm SI}\approx0.16\left(M/M_{\odot}\right)\;{\rm ms}=1.6\;\rm{ms}$. Also checking for trajectories with nonzero magnetic inclinations within the considered range of angles, we find that these values vary up to $\approx 5\%$.

We need to determine the limiting number density of particles $n$, which is consistent with the collisionless description. The general formula relating density $n$ to the mean free path $l_{\rm f}$, the cross section of relevant interaction $\sigma$, and the corresponding impact parameter $b$ reads 
\begin{equation}
 \label{dens}
n=\frac{1}{\sigma{}l_{\rm f}}=\frac{1}{\pi b^2 l_{\rm f}},
\end{equation}
where all quantities are in physical units (we omit the subscript $\rm{SI}$ for the rest of the discussion). As already commented on in Section\ \ref{intro}, we demand the particle mean free path to (at least) equal the characteristic length scale of the system. Therefore, we set $l_{\rm f}=(M/M_{\odot})\,1472\;\rm{m}\approx15\;\rm{km}$ for the black hole of ten solar masses. The interaction between particles is described as a classical Coulomb collision.

The analysis of elastic Coulomb scattering of two equal particles of mass $m$ and charge $\tilde{q}$ reveals \citep[e.g.,][]{jackson99} that the deflection angle $\chi$ of the incident particle with impact parameter $b$ and original velocity $v_{0}$ (measured in the laboratory frame in which the target particle is initially at rest) reads
\begin{equation}
 \label{deflection}
\cot\chi=\frac{2\pi\varepsilon_{0}mv_{0}^2 b}{\tilde{q}^2},
\end{equation}
 where $\varepsilon_{0}=8.85\times 10^{-12}\;\rm{F}\,\rm{m}^{-1}$ denotes the permittivity of the vacuum. The relative loss of kinetic energy of the incident particle (which is transferred to the target particle) $p\equiv (E_{\rm kin}^{0}-E_{\rm kin}^{\rm final})/E_{\rm kin}^{0}$ equals $p=\sin^2\chi$. The number density $n$ may thus be expressed as follows
\begin{equation}
 \label{dens2}
n=\frac{1}{\pi l_{\rm f}}\left[\frac{2\pi\varepsilon_0 m v_0^2}{\cot\left(\arcsin p^{1/2}\right)\tilde{q}^2} \right]^2.
\end{equation}
Setting $v_0=v_{\perp}=0.1c$, electron mass $m=9.11\times10^{-31}\;\rm{kg}$, electron charge $\tilde{q}=1.6\times10^{-19}\;\rm{C}$ and $l_{\rm f}=15\;\rm{km}$, we obtain the density $n$ as a function of $p$. If we allow the particle to lose maximally $1$ percent of its kinetic energy in a single collision and set $p=0.01$ (maximal deflection angle $\chi\approx6^{\circ}$), the resulting number density of electrons is $n\approx6.8\times 10^{17}\;\rm{m}^{-3}$.

The total power of radiation $P$ emitted by non-relativistic point charge $\tilde{q}$ with acceleration $a$ is given by Larmor formula \citep[e.g.,][]{rybicki79}
\begin{equation}
 \label{larmor}
P=\frac{\tilde{q}^2a^2}{6\pi\varepsilon_{0}c^3}.
\end{equation}
The special case of radial acceleration (perpendicular to the velocity) is usually denoted as cyclotron radiation. This is often produced by charges accelerated solely by the Lorentz force in the magnetic field, and in such a case one sets $a=v_{\perp}^2/r_{\rm{L}}$ in a Larmor formula. In our case, however, the particles are also accelerated gravitationally, and we obtain the actual acceleration $a$ by approximating our typical orbit by a trajectory of uniform circular motion and setting $a=2\pi v_{\perp}/T_{\varphi}$.

To obtain the radiation power of whole ensemble of $N$ electrons in the confinement we employ the dipole approximation \citep{rybicki79} which allows us to ignore differences in the retarded times of each particle as long as the typical size of the system $L=(M/M_{\odot})\,1472\;\rm{m}\approx15\;\rm{km}$, and the typical time scale of changes within the system, $\tau=T_{\varphi}=2.6\;\rm{ms}$, fulfills $\tau\gg L/c$. The overall radiation power is then given as $P_{\rm tot}=NP=nVP$. The volume of the confinement $V$ is estimated as the interior of the torus of inner radius $r=2.5\,(M/M_{\odot})\,1472\;\rm{m}$ and outer radius $R=3.5\,(M/M_{\odot})\,1472\;\rm{m}$ which reads $V\approx5\,(M/M_{\odot})^3\times 10^{10}\;\rm{m}^3=5\times 10^{13}\;\rm{m}^3$. 

Putting all the pieces together we obtain the total radiation power as 
\begin{equation}
 \label{powertot}
P_{\rm tot}=nVP\approx\left(\frac{q_{\rm e}}{q}\right)^2 \;\rm{Watt}.
\end{equation}

A counter-intuitive dependence on the specific charge of particles $q$ is due to the fact that we constrain the number density by only  considering Coulomb collisions which become less effective with decreased specific charges of colliding particles. The validity of the above formula is thus limited to the particles with $q\gg1$. The case of particles with lower $q$ would demand further discussion. We also note that although the volume of the confinement increases as $V\propto M^3$, the radiation power of a single particle decreases as $P\propto M^{-2}$ and the number density as $n\propto M^{-1}$. Thus, the total power $P_{\rm tot}$ is small, and it does not scale with the mass of the black hole $M$. 

Besides the radiation related to acceleration due to external fields, we may also estimate the power of bremsstrahlung (free--free radiation) of colliding particles \citep{jackson99} 
\begin{equation}
 \label{brem}
P_{\rm tot}^{\rm ff}=nVP^{\rm ff}=nV\frac{\tilde{q}^6}{\left(4\pi\varepsilon_{0}\right)^3}\frac{2}{3c^3}\frac{1}{m^2v_{0}b^3}\frac{v_{0}}{l}\approx\left(\frac{M}{10M_{\odot}}\right)\left(\frac{m}{m_{\rm e}}\right)^3\left(\frac{\tilde{q}_{\rm e}}{\tilde{q}}\right)^4 4.6\times10^{12}\;\rm{Watt},
\end{equation}
where we substituted the value of impact parameter $b$ which corresponds to the highest-allowed relative loss of kinetic energy $p=0.01$. A substantially higher value compared to (\ref{powertot}) looks more promising, nevertheless, in this case we do not expect the resulting radiation to encode dynamical frequencies of the orbit in PDS, and the chance of distinguishing dynamic regime of colliding particles is thus minimal.

\end{document}